\documentclass[a4paper,10pt,twocolumn]{article}
\usepackage{wallpaper}
\ThisULCornerWallPaper{1}{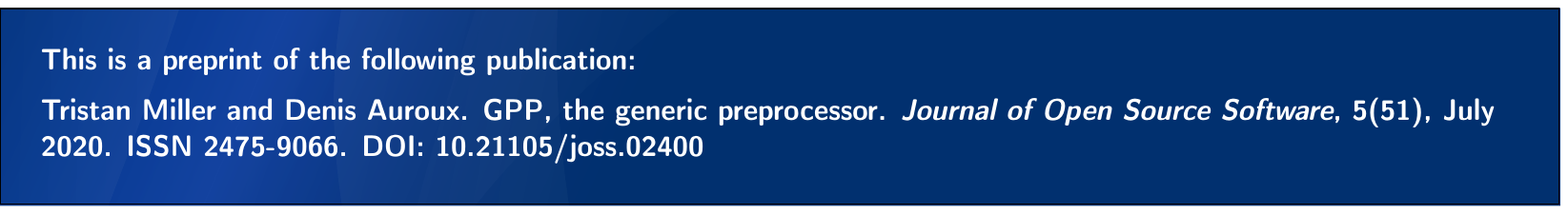}
\usepackage[canadian]{babel}
\usepackage{microtype}
\usepackage{url}
\usepackage[hidelinks]{hyperref}
\usepackage{csquotes}
\usepackage{xpatch}
\usepackage[%
backend=biber,
bibencoding=utf8,
style=authoryear-comp,
bibstyle=authoryear,
sorting=nyt,
maxcitenames=2,
uniquelist=false,
uniquename=init,
dashed=false,
giveninits=false,
isbn=true,
doi=true,
]{biblatex}
\xpatchbibmacro{volume+number+eid}{%
  \setunit*{\adddot}%
}{%
}{}{}

\DeclareFieldFormat[article,periodical]{number}{\mkbibparens{#1}}
\DeclareFieldFormat[article,periodical]{pages}{#1}
\DeclareFieldFormat[article,inproceedings,incollection,inbook]{title}{#1} 
\renewbibmacro{in:}{}

\hypersetup{
  pdftitle={GPP, the Generic Preprocessor},
  pdfauthor={Tristan Miller and Denis Auroux},
}

\begin{document}

  \title{GPP, the Generic Preprocessor}
\author{Tristan Miller\\
Austrian Research Institute for Artificial Intelligence\\
Freyung 6/3, 1010 Vienna, Austria\\
ORCID: 0000-0002-0749-1100
\and
Denis Auroux\\
Department of Mathematics, Harvard University\\
1 Oxford Street, Cambridge, MA 02138, USA}
\date{}

\maketitle

\section*{Summary}

In computer science, a \emph{preprocessor} (or \emph{macro processor}) is a tool that programatically alters its input, typically on the basis of inline annotations, to produce data that serves as input for another program.  Preprocessors are used in software development and document processing workflows to translate or extend programming or markup languages, as well as for conditional or pattern-based generation of source code and text.  Early preprocessors were relatively simple string replacement tools that were tied to specific programming languages and application domains, and while these have since given rise to more powerful, general-purpose tools, these often require the user to learn and use complex macro languages with their own syntactic conventions.  In this paper, we present GPP, an extensible, general-purpose preprocessor whose principal advantage is that its syntax and behaviour can be customized to suit any given preprocessing task.  This makes GPP of particular benefit to research applications, where it can be easily adapted for use with novel markup, programming, and control languages.

\section*{Background}

Preprocessors date back to the mid-1950s, when they were used to extend individual assembly languages with constructs that would later be found in high-level programming languages~\autocite{layzell1985history}.  These languages, in turn, fostered the development of yet more special-purpose preprocessors aimed at providing even higher-level constructs, such as conditional loops and other control structures in FORTRAN~\autocite{meissner1975extending} and COBOL~\autocite{triance1980structured}.  The need for generalized, language-independent tools was eventually recognized~\autocite{mcilroy1960macro}, leading to the development of general-purpose preprocessors such as GPM~\autocite{strachey1965general} and ML/I~\autocite{brown1967mli}.

By the end of the 1960s, preprocessors had attracted a considerable amount of attention, by computing theorists and practitioners alike, and their use in software engineering had expanded beyond the augmentation and adaptation of programming languages. A survey paper by \textcite{brown1969survey} identified four broad application areas: language extension, systematic searching and editing of source code, translation between programming languages, and code generation (i.e., simplifying the writing of highly repetitive code, parameterizing a program by substituting compile-time constants, or producing variants of a program by conditionally including certain statements or modules).  While the first three of these application areas have largely been rendered obsolete by today's integrated development environments and expressive, feature-rich programming languages, implementing software variability with language-specific and general-purpose preprocessors remains commonplace~\autocite{kaestner2012type,apel2013feature}.

Text processing became another main application area for preprocessors, in particular to generate documents on the basis of user-specified conditions or patterns, and to convert between document markup languages~\autocite{walden2014macro}.  The earliest such uses were ad-hoc repurposings of programming language--specific preprocessors to operate on human-readable texts~\autocite{keese1964note,stallman2020c}; these were soon supplanted by text-specific macro languages such as TRAC~\autocite{mooers1965trac}, which were positioned as tools for stenographers and other writing professionals.  More recently it has been common to use general-purpose preprocessors~\autocite{pesch1992configurable,mailund2019introducing}.

\section*{Statement of Need}

Criticism of preprocessors commonly focuses on the idiosyncratic languages they employ for their own built-in directives and for users to define and invoke macros.  The languages of early preprocessors were derided as ``clumsy and restrictive''~\autocite{layzell1985history} and ``hard to read''~\autocite{brown1969survey}, and even modern preprocessors are sometimes attacked for relying on ``the clumsiness of a separate language of limited expressiveness''~\autocite{ernst2002empirical} or, at the other extreme, for being overly complicated, quirky, opaque, or hard to learn, even for experienced programmers and markup users~\autocite{pesch1992configurable,paddon1993shake,ernst2002empirical}.

Our general-purpose preprocessor, GPP, avoids these issues by providing a light\-weight but flexible macro language whose syntax can be customized by the user.  The tool's built-in presets allow its directives to be made to resemble those of many popular languages, including HTML and \TeX.  This greatly reduces the learning curve for GPP when it is used with these languages, eliminates the cognitive burden of repeatedly ``mode switching'' between source and preprocessor syntax when reading or composing, and allows existing syntax highlighters and other tools to process GPP directives with little or no further configuration.  Furthermore, users are not limited to using these presets, but can fully define their own syntax for GPP directives and macros. This makes GPP particularly attractive for use in research and development, where its syntax can be readily adapted to match novel programming and markup languages.

GPP's independence from any one programming or markup language makes it more versatile than the C Preprocessor, which was formerly ``abused'' as a general text processor and is still sometimes (inappropriately) used for non-C applications~\autocite{stallman2020c}.  While GPP is less powerful than m4~\autocite{seindal2016m4}, it is arguably more flexible, and supports all the basic operations expected of a modern, high-level preprocessing system, including conditional tests, arithmetic evaluation, and POSIX-style wildcard matching (``globbing'').  In addition to macros, GPP understands comments and strings, whose syntax and behaviour can also be widely customized to fit any particular purpose.

\section*{GPP in research}

GPP has already been integrated into a number of third-party projects in basic and applied research.  These include the following:
\begin{itemize}
\item The Waveform Definition Language (WDL) is Caltech Optical Observatories' C-like language for programming astronomical research cameras.  WDL uses GPP to preprocess configuration files containing signals and parameters specific to the camera controllers, flags setting the devices' operating modes and image properties, and timing rules.  According to the developers, GPP was chosen over the C Preprocessor ``for added flexibility and to avoid some C-like limitations''~\autocite{kaye2017waveform}.
\item XSB is a research-oriented, commercial-grade logic programming system and Prolog compiler.  The developers chose to make GPP XSB's default preprocessor because it ``maintains a high degree of compatibility with the C preprocessor, but is more suitable for processing Prolog programs''~\autocite{swift2017xsb}.
\item C-Control Pro is a family of electronic microcontrollers produced by Conrad Electronic; they are specifically designed for industrial and automotive applications. The official software development kit includes a modified version of GPP for use with the products' BASIC and Compact-C programming languages~\autocite{schirm2007messen}.%
\item SUS is a tool that allows system administrators to exercise fine-grained control over how users can run commands with elevated privileges.  It has a sophisticated control file syntax that is preprocessed with GPP~\autocite{gray2001sus}.
\end{itemize}
Apart from these uses, GPP is occasionally cited as previous or related work in scholarly publications on metaprogramming or compile-time variability of software~\autocite{baxter2001preprocessor,lotoreychik2006metaprogrammirovaniye,dreiling2010feature,blendinger2010filesystem,kaestner2012type,apel2013feature,zmiry2016lola,behringer2017projectional}.

\section*{Acknowledgments}

Tristan Miller is supported by the Austrian Science Fund~(FWF) under project M\,2625-N31. Denis Auroux is partially supported by NSF grant DMS-1937869 and by Simons Foundation grant \#385573. The Austrian Research Institute for Artificial Intelligence is supported by the Austrian Federal Ministry for Science, Research and Economy.

\printbibliography

\end{document}